\documentclass[11pt]{article}

\usepackage[linesnumbered,lined,boxed,commentsnumbered]{algorithm2e}

\usepackage{arxiv}

\usepackage[utf8]{inputenc}
\usepackage{epsfig}
\usepackage{graphicx}
\usepackage{ulem}
\usepackage[letterpaper]{geometry}
\usepackage{multirow}
\usepackage{graphicx}
\usepackage[usenames,dvipsnames]{color}
\usepackage{todonotes}
\usepackage{bbm}

\newcommand{\E}{\mathbb E}

\usepackage[numbers]{natbib}
\setcitestyle{aysep={}}

\usepackage{graphicx}
\usepackage[colorlinks=true, allcolors=blue]{hyperref}

\usepackage{paralist}
\usepackage{amsmath}
\usepackage{amssymb}
\usepackage{amsthm}

\usepackage{tikz}
\usetikzlibrary{arrows}
\usetikzlibrary{arrows.meta}

\usepackage[]{todonotes}   
\usepackage{soul}

\author{
Bogumi\l{} Kami\'nski\thanks{Decision Analysis and Support Unit, SGH Warsaw School of Economics, Warsaw, Poland; \texttt{bkamins@sgh.waw.pl}}
\And
Tomasz Olczak\thanks{Decision Analysis and Support Unit, SGH Warsaw School of Economics, Warsaw, Poland; \texttt{tolczak@gmail.com}}
\And
Bartosz Pankratz\thanks{Decision Analysis and Support Unit, SGH Warsaw School of Economics, Warsaw, Poland; \texttt{bartosz.pankratz@ryerson.ca}}
\And
Pawe\l{}~Pra\l{}at\thanks{Department of Mathematics, Toronto Metropolitan University, Toronto, ON, Canada; \texttt{pralat@ryerson.ca}}
\And
Fran\c{c}ois Th\'eberge\thanks{Tutte Institute for Mathematics and Computing, Ottawa, ON, Canada; \texttt{theberge@ieee.org}}
}

\title{Properties and Performance of the ABCD${e}$ Random Graph Model with Community Structure}

\begin{document}

\maketitle

\setcounter{footnote}{0}

\begin{abstract}
In this paper, we investigate properties and performance of synthetic random graph models with a built-in community structure. Such models are important for evaluating and tuning community detection algorithms that are unsupervised by nature.
We propose \textbf{ABCDe}---a multi-threaded implementation of the \textbf{ABCD} (Artificial Benchmark for Community Detection) graph generator.
We discuss the implementation details of the algorithm and compare it with both the previously available sequential version of the \textbf{ABCD} model and with the parallel implementation of the standard and extensively used \textbf{LFR} (Lancichinetti--Fortunato--Radicchi) generator. We show that \textbf{ABCDe} is more than ten times faster and scales better than the parallel implementation of \textbf{LFR} provided in \texttt{NetworKit}. Moreover, the algorithm is not only faster but random graphs generated by \textbf{ABCD} have similar properties to the ones generated by the original \textbf{LFR} algorithm, while the parallelized \texttt{NetworKit} implementation of \textbf{LFR} produces graphs that have noticeably different characteristics.
\end{abstract}

\section{Introduction}

The standard and extensively used method for generating artificial networks that have community structure is the \textbf{LFR} (Lancichinetti--Fortunato--Radicchi) graph generator~\cite{lfr}. Despite the fact that this is clearly a very good model, it is known to have some scalability limitations and it is challenging to analyze it theoretically. Moreover, the mixing parameter $\mu$, the main parameter of the model guiding the strength of the communities, has a non-obvious interpretation and so can lead to unnaturally-defined networks, see~\cite{ABCD} for a detailed discussion.

An alternative random graph model with community structure and power-law distribution for both degrees and community sizes is the \textbf{A}rtificial \textbf{B}enchmark for \textbf{C}ommunity \textbf{D}etection graph (\textbf{ABCD}). In~\cite{ABCD} it is shown that the new model is fast, simple, and can be easily tuned to allow the user to make a smooth transition between the two extremes: pure (disjoint) communities and random graph with no community structure. Moreover, in~\cite{ABCD-theory} the modularity function of \textbf{ABCD} is theoretically analyzed and it is confirmed that its asymptotic behaviour is consistent with simulations on smaller experimental graphs. (The modularity function is, arguably, the most important graph property of networks in the context of community detection. Indeed, the modularity function is often used to measure the presence of community structure in networks. It is also used as a quality function in many community detection algorithms, including the widely used Louvain algorithm~\cite{Louvain}.) On the other hand, because of its similarity to \textbf{LFR}, \textbf{ABCD} can be expected to preserve most of its natural graph properties and parameters. We verify this claim in this paper using simulation. Hence, \textbf{ABCD} (or \textbf{ABCDe} introduced and discussed in this paper) may successfully replace the \textbf{LFR} generator when scalability becomes a bottleneck.

In this paper we introduce the implementation of the \textbf{ABCD} generator that uses multiple threads for processing, \textbf{ABCDe}. The goal of parallelizing sequential \textbf{ABCD} is to speed up computations and thus allow for handling of larger graphs (having more nodes or being more dense).
We describe the challenges of parallelization of this algorithm and the approach we took to ensure both its performance and reproducibility of generated graphs.
We analyze the \textbf{ABCDe} properties using simulation, taking into account two important aspects. First, we analyze the properties of the graphs that it generates and compare them to graphs generated by the \textbf{LFR} algorithm under matching parameterization. Next, we analyze the speed of \textbf{ABCDe} against the single threaded (sequential) implementation of \textbf{ABCD} and selected implementations of the \textbf{LFR} generator. The two selected implementations are: the original \textbf{LFR} algorithm~\cite{lfr} and its fast implementation, that optionally uses multiple threads, provided in the \texttt{NetworKit} package~\cite{networkit}\footnote{\url{https://networkit.github.io/index.html}}.

\section{The Design of the {\bf ABCDe} Generator}

\subsection{Building Blocks of the Algorithm}

As a preliminary information let us start with introducing two building blocks of the {\bf ABCD} model: configuration model and Chung-Lu model. Let $\textbf{w}=(w_1, \ldots, w_n)$ be any vector of $n$ non-negative integers. Our goal is to be able to build two types of random graphs on $n$ nodes, the first one will have a given degree sequence $\textbf{w}$ (configuration model) and the second one will only have the expected degree sequence $\textbf{w}$ (Chung-Lu model).

A random multi-graph $\mathcal{M}(\textbf{w})$ with a given degree sequence known as the \textbf{configuration model} (sometimes called the \textbf{pairing model}) was first introduced by Bollob\'{a}s~\cite{CM}. Assuming that $W:=\sum_{i=1}^n w_i$ is even, let us consider $W$ points partitioned into $n$ labelled buckets $v_1,\ldots,v_n$; bucket $v_i$ consists of $w_i$ points. A \textbf{pairing} of these points is a perfect matching into $W/2$ pairs.  (There are $W! / ((W/2)! 2^W)$ such pairings.) Given a pairing $P$, we may construct a multi-graph $G(P)$, with loops and parallel edges allowed, as follows: the nodes are the buckets $v_1,\ldots, v_n$, and a pair $\{x,y\}$ in $P$ corresponds to an edge $\{v_i,v_j\}$ in $G(P)$ if $x$ and $y$ are contained in the buckets $v_i$ and $v_j$, respectively. We take a pairing $P$ uniformly at random from the family of all pairings of $W$ points and set $\mathcal{M}(\textbf{w}) = G(P)$.

In the Chung-Lu model~\cite{CL} that generates graph $\mathcal{C}(\mathbf{w})$ on the node set $[n] = \{1, \ldots, n\}$, each set $e=\{i,j\}$, $i,j \in [n]$, is independently sampled as an edge with  probability given by:
\[
\Pr(i,j) =
\begin{cases}
\frac{w_i w_j}{W}, & i \ne j \\
\frac{(w_i)^2}{2W}, & i = j.
\end{cases}
\]
(Let us mention about one technical assumption. Note that it might happen that $\Pr(i,j)$ is greater than one and so it should really be regarded as the expected number of edges between $i$ and $j$; for example, as suggested in Newman~\cite{Newman_book}, one can introduce a Poisson-distributed number of edges with mean $\Pr(i,j)$ between each pair of nodes $i$, $j$. However, since typically the maximum degree $\Delta$ satisfies $\Delta^2 \le 2 |E|$ it rarely creates a problem and so we may assume that $\Pr(i,j) \le 1$ for all pairs.)

One desired property of this random model is that it yields a distribution that preserves the expected degree for each node, namely: for any $i \in [n]$,
$$
\E[\deg(i)] = \sum_{j \in [n] \setminus \{i\}} \frac{w_i w_j}{W} + 2 \cdot \frac{(w_i)^2}{2W} = \frac{w_i}{W} \sum_{j \in [n]} w_j = w_i.
$$

In summary, both models are similar. The difference between them is that \textbf{configuration model} ensures that the required node degree sequence is reproduced exactly, while \textbf{Chung-Lu model} produces this degree sequence in expectation.

\subsection{Structure of the {\bf ABCD} Graph}

In this section, we briefly discuss the {\bf ABCD} models; details can be found in~\cite{ABCD} or in~\cite{ABCD-theory}.
As in {\bf LFR}, for a given number of nodes $n$, we start by generating a power law distribution both for the degrees and community sizes. Those are governed by the power law exponent parameters $(\gamma,\beta)$. We also provide additional information to the model, again as it is done in {\bf LFR}, namely, the average and the maximum degree, and the range for the community sizes. The user may alternatively provide a specific degree distribution and/or community sizes.

For each community, we generate a random {\it community} subgraph on the nodes from a given community using either the \textbf{configuration model} (see~\cite{CM}) which preserves the exact degree distribution, or the \textbf{Chung-Lu model} (see~\cite{CL}) which preserves the expected degree distribution.
On top of it, we independently generate a {\it background} random graph on all the nodes that is generated the same way as the community graphs. Everything is tuned properly so that the degree distribution of the union of all graphs follows the desired degree distribution (only in expectation in the case of the Chung-Lu variant).
In particular, the mixing parameter $\xi$ guides the proportion of edges which are generated via the background graph. In the two extreme cases, when $\xi=1$ the graph has no community structure while if $\xi=0$, then we get disjoint communities.
In order to generate simple graphs, we may have to do some re-sampling or edge re-wiring, which as described in~\cite{ABCD}. This two-step process is similar to the highly scalable {\bf BTER} model~\cite{bter}. (Similarly to {\bf LFR} and {\bf ABCD}, {\bf BTER} generates graphs with a given degree distribution but the main objective is different: it aims to preserve per-degree clustering coefficients. In particular, this model does not have the same type of community structure as in {\bf LFR} and {\bf ABCD}.)

During this process, larger communities will additionally get some more internal edges due to the background graph. As argued in~\cite{ABCD}, this ``global'' variant of the model is more natural and so we recommend it. However, in order to provide a variant where the expected proportion of internal edges is exactly the same for every community (as it is done in {\bf LFR}), we also provide a ``local'' variant of {\bf ABCD} in which the mixing parameter $\xi$ is automatically adjusted for every community. Both variants preserve the same number of edges between communities. The difference is how the degree of each node is split into internal and external degree. The {\bf LFR} model, as well as our local variant of the {\bf ABCD} model, keep the same fraction of neighbours to be internal neighbours for all nodes, regardless how large the community this node belongs to is. As a result, small communities become much denser than large communities. On the other hand, in the global variant of the {\bf ABCD} model, the internal degree naturally depends on the size of the associated community.

\begin{figure}
\centering
\includegraphics[scale=0.6]{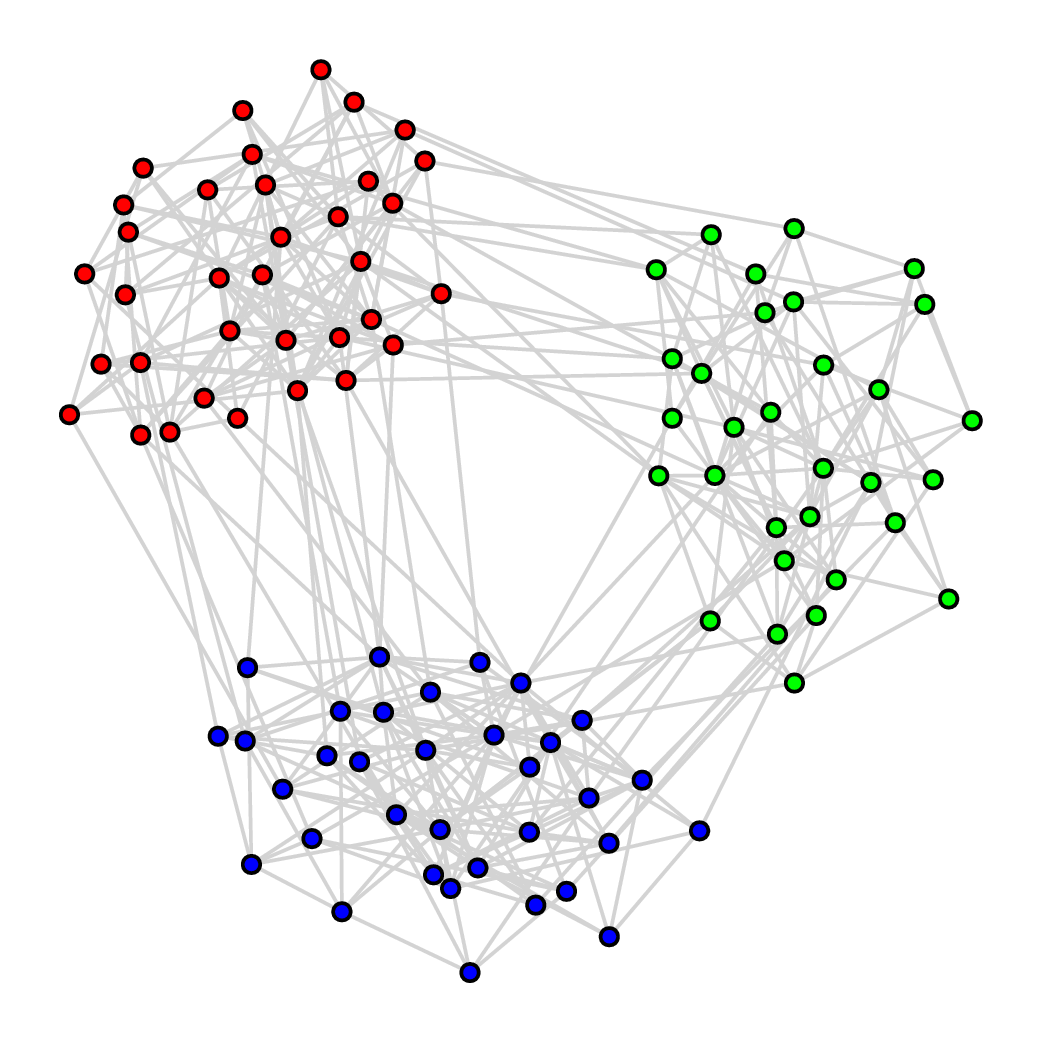}
\hspace{.5cm}
\includegraphics[scale=0.6]{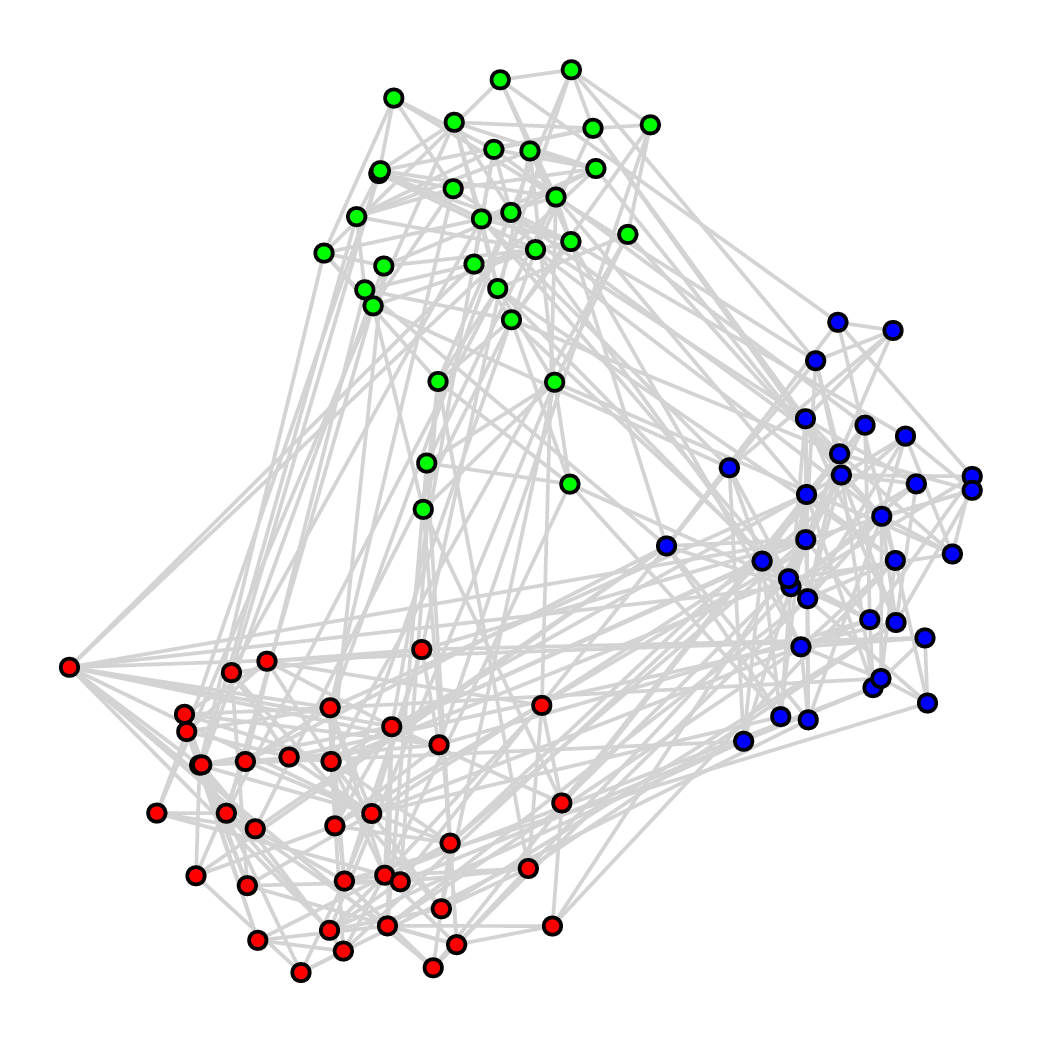}
\caption{Two examples of \textbf{ABCD} graphs with low level of noise ($\xi=0.2$, left) and large level of noise ($\xi=0.4$, right).}
\label{fig:examples}
\end{figure}

Two examples of \textbf{ABCD} graphs on $n=100$ nodes are presented in Figure~\ref{fig:examples}.
Degree distribution was generated with power law exponent $\gamma=2.5$ with minimum and maximum values 5 and 15, respectively. Community sizes were generated with power law exponent $\beta = 1.5$ with minimum and maximum values 30 and 50, respectively; communities are shown in different colours. The global variant and the configuration model was used to generate the graphs. The left plot has the mixing parameter set $\xi=0.2$ while the ``noisier'' graph on the right plot has the parameter fixed to $\xi=0.4$.

In this paper, we compare both ``global'' and ``local'' variants of the \textbf{ABCD} model (using the configuration model to generate communities as well as the background graph) against the classical \textbf{LFR} model.

\subsection{Approach to Parallelization of the {\bf ABCD} Generator}

The original {\bf ABCD} model was implemented using a sequential algorithm (Algorithm~\ref{alg:ABCD}). In~\cite{ABCD}, we discuss theoretical complexity of \textbf{ABCD} and \textbf{LFR} for this scenario.

Let us now switch to issues of parallelization of the process of generating \textbf{ABCD} graphs.
The model, being a union of community graphs and the background graph, is naturally structured for concurrent processing.
Since community graphs are disjoint, they can be generated independently of each other. Typically they are numerous and their individual sizes are relatively small when compared to the whole graph. Taken together they constitute a set of sufficiently granular tasks for efficient parallel processing.
Hence, in \textbf{ABCDe} (a multi-threaded version of \textbf{ABCD} generator) we distribute community graphs for parallel generation among the available CPU threads.

The main challenge for the design of parallelized \textbf{ABCDe} algorithm is posed by generation of the background graph. In the sequential version of the algorithm, the background graph is generated in the final stage of processing, after all the community graphs are generated. This order of processing enables detection of potential collisions, that is, duplicate edges in the background graph and one of the community graphs.
Since the background graph shares nodes with all the community graphs such collisions may arise and must be avoided to ensure the final graph to be simple.

Preserving the same processing sequence in the parallel algorithm would have deteriorating effect on performance as often the background graph is chosen to be so large that its generation consumes a non-negligible fraction of the overall processing time. Postponing generation of the background graph only after all the community graphs are generated would significantly increase the fraction of time spend in sequential processing in overall processing time thus limiting speedup from parallel processing as governed by Amdahl's law~\cite{Amd67}.

To work around this serialization bottleneck, one could parallelize generation of the background graph itself. Unfortunately, this approach is challenging by nature of a single graph generation algorithm.
Recall that we defined the \textbf{configuration model} graph by $G(P) = \mathcal{M}(\textbf{w})$ where
a pairing $P$ is sampled uniformly at random from the family of all pairings of $W:=\sum_{i=1}^n w_i$ points. This sampling is implemented in the following way. First, a random permutation $p := \{1, \ldots, m\} \mapsto \{1, \ldots, m\}$ where $m := W$ is generated using Fisher--Yates shuffle~\cite{FY43}. Then, a pairing $P_p := (p_1, \ldots, p_\frac{m}{2}) \mapsto (p_{\frac{m}{2}+1}, \ldots, p_m)$ is obtained and a multi-graph $G(P_p)$ is produced.

Fisher--Yates algorithm~\cite{FY43} is inherently sequential and its potential parallelization would involve high performance penalty due to thread synchronization thus cancelling gains from parallelization.
Therefore, in \textbf{ABCDe} the second best approach was taken where, in contrast to sequential \textbf{ABCD}, the background graph is generated \emph{in parallel} with community graphs.
To enable this change, generation of the background graph was split into two phases. In the first phase, composed of tasks independent of community graph generation, the disjoint background graph is generated and any internal self-loops or parallel edges rewired to ensure the graph to be simple (steps 6 and 7 in Algorithm~\ref{alg:ABCDe}). Note that this phase is indistinguishable from generation of community graphs. In the second phase, after all the community graphs \emph{and} the background graph were generated, individual graphs are merged to form the final \textbf{ABCDe} graph and any parallel edges resulting from the merger are rewired, if present (steps 9 and 10 in Algorithm~\ref{alg:ABCDe}).

To ensure well-balanced distribution of work among available threads and to minimize total processing time, we apply the following task allocation policy. Typically, the background graph is the largest one in the ensemble and its generation is the most time consuming. It is then the first graph picked up for processing by the first available thread so its generation could start as soon as possible. The remaining set of community graphs is placed in a random order into a FIFO queue and consumed for processing by a pool of available threads following a classical producer-consumer pattern. This generates a uniform workload in expectation. Note that this policy does not prevent unintended serialization in case of an extremely large background graph, as it would still be processed by a single thread only. In this case benefits of parallelization could be limited. Fortunately, this happens only for $\xi$ close to $1$, which is not normally used in practical applications (usually, $\xi$ less than $0.5$ is used).

Let us also make a short remark on the optimal number of threads. \textbf{ABCDe} will use \emph{all} the threads available for Julia process and the number of Julia threads can be specified when Julia process is started.
As a rule of thumb, for optimal \textbf{ABCDe} performance one should use the number of threads equal to the number of \emph{physical} (not logical) cores as \textbf{ABCDe} is a memory intensive algorithm. Increasing number of threads beyond this point will usually not make noticeable improvement or can even deteriorate performance as excess threads would be forced to share the same memory bus.

Finally, a note should be taken on reproducibility of results. The challenge is that generation of each graph is an independent task and both the order and thread allocation is unspecified and decided only at run-time by the operating system scheduler. This would lead to a situation where running the algorithm twice for the same parameterization could potentially produce different graphs. In order to solve this issue, each task is associated with a random number generator seed before any graph generation is started. This way the stream of pseudo-random numbers is ensured to be reproducible independently of task processing order, number of threads, and task to thread allocation.

In summary, we parallelize generation of the \textbf{ABCD} graph in the following way:
\begin{enumerate}
    \item [Step 1:] sequentially assign nodes to communities and split their degrees into community and background graph;
    \item [Step 2:] sequentially put all community graphs and background graph in a queue of tasks; assign a random number generator seed to each task;
    \item [Step 3:] using parallel map algorithm, generate all community graphs and background graph independently;
    \item [Step 4:] sequentially create a union of all graphs;
    \item [Step 5:] sequentially resolve all conflicts between the background graph and community graphs to ensure that the resulting graph is simple.
\end{enumerate}

Steps 1, 2, 4, and 5 are done sequentially. Note that steps 1, 2, and 4 are computationally cheap so that there would be no noticeable benefit of running them using a parallel algorithm. Step 5 could, in general, be potentially expensive. Fortunately, for large and sparse graphs the number of conflicts that need to be resolved in this step is small and does not significantly affect the overall run-time of the algorithm. Step 3, the most expensive part of the algorithm, is executed using multi-threading. In our experiments, we verify how increasing number of threads affects the run-time of the whole algorithm.

In general, the computational complexity of the \textbf{ABCDe} algorithm is linear in the number of edges generated, similarly to the sequential \textbf{ABCD} algorithm, as discussed in~\cite{ABCD}. However, as it will be seen in Section~\ref{sec:speedtest}, the scaling of the algorithm with increasing $n$ has a growing constant factor. The reason for this behaviour is twofold. The first is theoretical---increasing $n$ changes degre distribution and community size distribution of the generated graph. These changes lead to changes of cost of collision resolution process. The second is technical---as graph size increases, the number of operations that are performed in CPU cache decreases. For large graphs most operations end up to be cache misses. Since reading data from CPU cache is much faster than fetching data from a new area of RAM, again the constant factor in the algorithm increases.

\begin{algorithm}[H]
\caption{\textbf{ABCD} (sequential)}\label{alg:ABCD}
\KwIn{vector of node degrees}
\KwOut{list of edges}
\For{node $\in$ nodes}{
    assign node to a community\;
    split the degree among the community and the background graph\;
}
\For{community $\in$ communities}{
    generate a community graph\;
    rewire self-loops and parallel edges to ensure the community graph is simple\;
}
generate the background graph\;
rewire self-loops and parallel edges to ensure the background graph is simple\;
create union of community graphs and the background graph\;
rewire parallel edges to ensure the output graph is simple\;
\Return edges\;
\end{algorithm}

\begin{algorithm}[H]
\caption{\textbf{ABCDe} (parallel)}\label{alg:ABCDe}
\SetAlgoLined
\KwIn{vector of node degrees}
\KwOut{list of edges}
\For{node $\in$ nodes}{
    assign node to a community\;
    split the degree among the community and the background graph\;
}
\ForPar{graph $\in$ community graphs $\cup$ \{background graph\} }{
    generate a graph\;
    rewire self-loops and parallel edges to ensure the graph is simple\;
}
create union of community graphs and the background graph\;
rewire parallel edges to ensure the output graph is simple\;
\Return edges\;
\end{algorithm}

\section{Properties of the Generated Graphs}

In this section, we present the results of several experiments comparing the properties of graphs generated using the sequential \textbf{ABCD} model and parallelized \textbf{ABCDe} model (both global and local variants) against the original \textbf{LFR} implementation~\cite{lfr} as well as its fast \texttt{NetworKit} implementation~\cite{networkit}. Note that the \textbf{ABCD} and \textbf{ABCDe} local variant of our algorithm was developed with the goal to reproduce the behaviour of the original \textbf{LFR} implementation. On the other hand, the \textbf{ABCD} and \textbf{ABCDe} global variant was proposed as an alternative to overcome some properties of the the original \textbf{LFR} implementation that we believe are not desirable---see~\cite{ABCD} for more details.

\subsection{Experiment Setup}

Properties were tested on graph with $n=10{,}000$ nodes. Graphs were generated with various values of the mixing parameter $\xi$ (namely, values between 0 and 1 with a step equal to $0.05$). Recall that the parameter $\xi$ is the main parameter of the model, responsible for the level of noise. The detailed results can be found in the accompanying notebook; we present the result for a specific value of $\xi=0.5$ below. Both variants of the \textbf{ABCD} model were tested for a given value of $\xi$ but in the case of the two variants of the \textbf{LFR} model, the value of its parameter $\mu$, the counterpart of $\xi$ in the \textbf{ABCD} model, was computed using the following formula:
\begin{equation}
\label{mu_approx}
    \mu = \xi \left( 1 - \sum_{\ell \in [k]} \frac{W_{\ell}}{W} \right),
\end{equation}
where $W$ is the volume of $G$ and $W_{\ell}$ is the volume of nodes that belong to $\ell$'th community.

In order to increase the comparability of all algorithms, we used the following coupling. For a given set of parameters, we pre-generated the degree distributions and the community sizes. Such pre-generated sequences were used in all four models. In both cases, the distributions were sampled using the discrete power-law distribution with truncation range. The samplers included in the package with the \textbf{ABCD} algorithm were used to do this.

As with the parameter $\xi$, various exponents of the power-law distributions were tested (namely, $\beta \in \{1.1, 1.5, 1.9\}$ and $\gamma \in \{2.1, 2.5, 2.9\}$) and details can be found in the associated notebook. However, in the figures we present, the degree distributions were generated with the exponent $\gamma = 2.5$, the minimum degree $\delta$ equal to 5, and the maximum degree equal to $\sqrt{n}$. Community sizes were generated with the exponent $\beta = 1.5$ and the lower and the upper bounds for the sizes of communities were functions of $n$, namely, there were equal to $0.005n$ and $0.2n$, respectively.

In the figures we present properties of both sequential \textbf{ABCD} algorithm and the \textbf{ABCDe} algorithm run with 32 threads to show that indeed, as expected from the description of the algorithms, the properties of the produced graphs are the same.

\subsection{Results}

For every sweep of the parameters of the model, 30 graphs were generated and for each of them 10 representative graph properties were investigated.
In the remaining of this section, we summarize the main findings based on the experiments. We will discuss each parameter independently but there are some general observations that we noticed.
Given the results of the measurements of 10 different graph characteristics, one can observe that both \textbf{ABCD} variants are much more similar to the original \textbf{LFR} implementation than the \texttt{NetworKit} implementation of \textbf{LFR}. Not surprisingly, the local variant of the \textbf{ABCD} model is especially close to the original \textbf{LFR} model.

Since the evaluated graph parameters are standard and well-known, we do not formally define them. We direct the reader to any book on mining complex networks such as~\cite{kaminski2021mining}.

\subsubsection{Clustering Coefficients}

The global clustering coefficient of a given graph $G$ is the ratio of three times the number of triangles to the number of pairs of adjacent edges. This parameter has a nice and important interpretation: given a random pair of adjacent edges, the global clustering coefficient is the probability that those three nodes form a triangle. As a result, it is often used as a measure of the presence of the so-called triadic closure, a natural mechanism present in many complex networks. For example, in a social network, strong triadic closure occurs because there is an increased opportunity for nodes $x$ and $z$ with common neighbour $y$ to meet.

The local clustering coefficient is defined for each node $v \in V(G)$ as the number of triangles this node is part of, divided by the total number of distinct pairs of neighbouring nodes for $v$. In other words, it is the probability that two random neighbours of $v$ are adjacent. The average local clustering coefficient of $G$ is obtained by averaging the local clustering coefficient over all nodes $v \in V(G)$.
Formulas for those coefficients can be found in Chapter~1 of~\cite{kaminski2021mining}.

\begin{figure}
\centering
\includegraphics[scale=0.48]{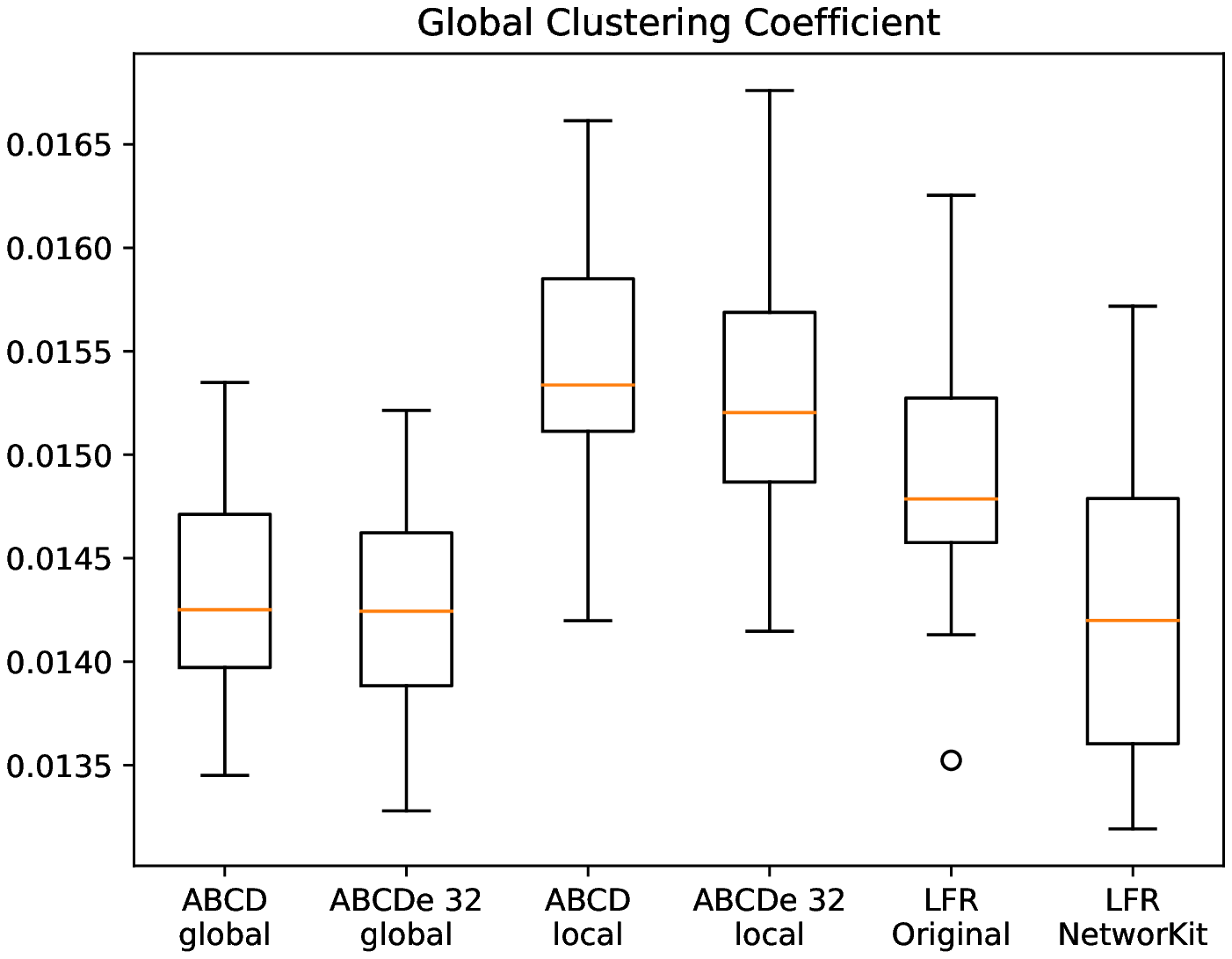}
\includegraphics[scale=0.48]{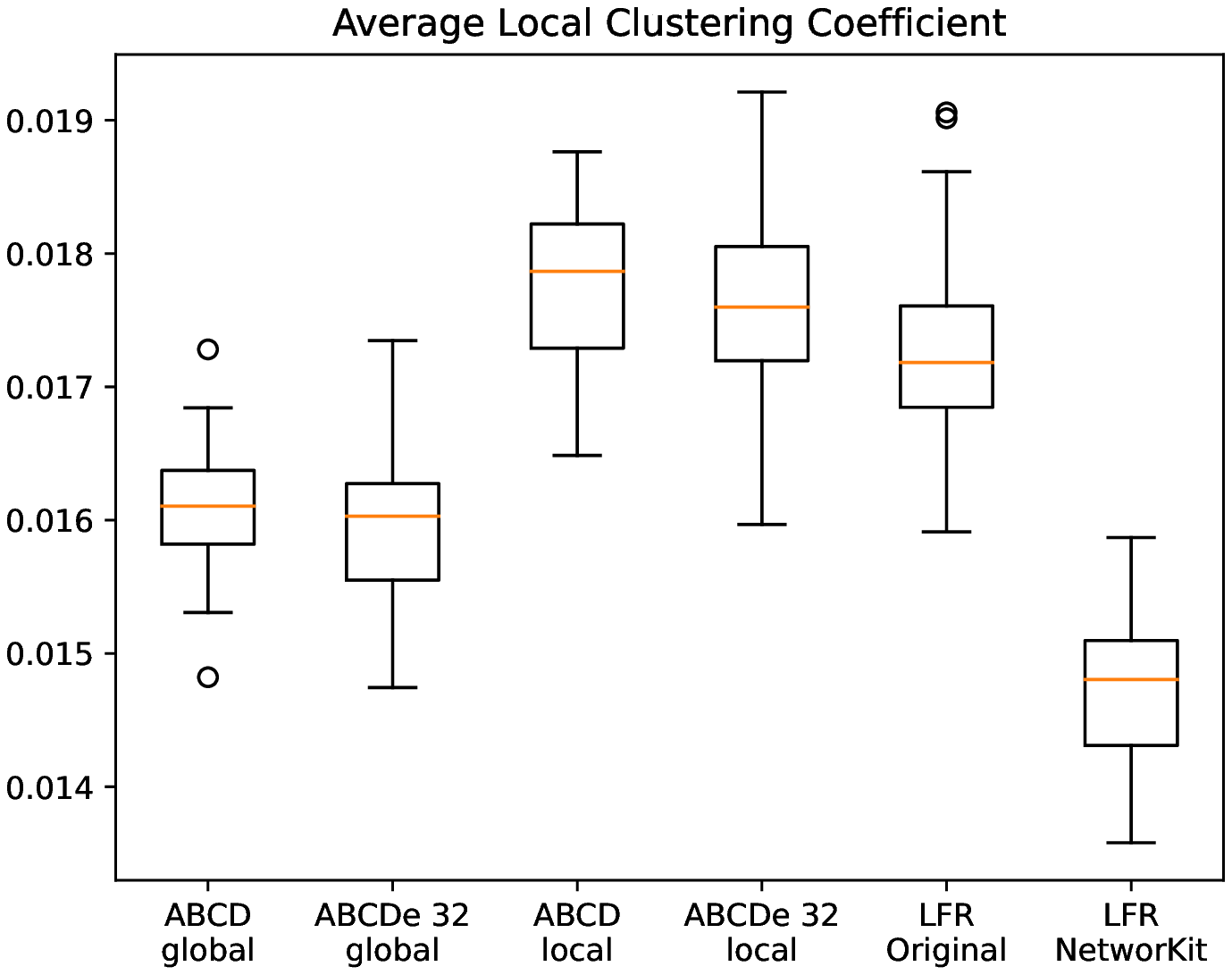}
\caption{Comparison of the distribution of the global clustering coefficient (left) and average local clustering coefficient (right).}\label{fig:cc}
\end{figure}

In Figure~\ref{fig:cc}, we see that the global clustering coefficient is similar for all four compared algorithms. On the other hand, the average local clustering coefficient is most similar between the local variant of \textbf{ABCD} and the original \textbf{LFR}, while \texttt{NetworKit} \textbf{LFR} differs the most. Havins said that, the observed differences are not substantial. Finally, let us note that the corresponding numerical values (for both variants) are rather low, comparable to what one would expect from a random chance based on the global density of the corresponding graphs. This is not surprising as none of these models aim to produce graphs with large clustering coefficients, a typical property of the so-called small-world networks. One example of a random graph with large clustering coefficient is the classical Watts and Strogatz model~\cite{Watts-Strogatz}.

\subsubsection{Node Centralities}

\begin{figure}
\centering
\includegraphics[scale=0.3]{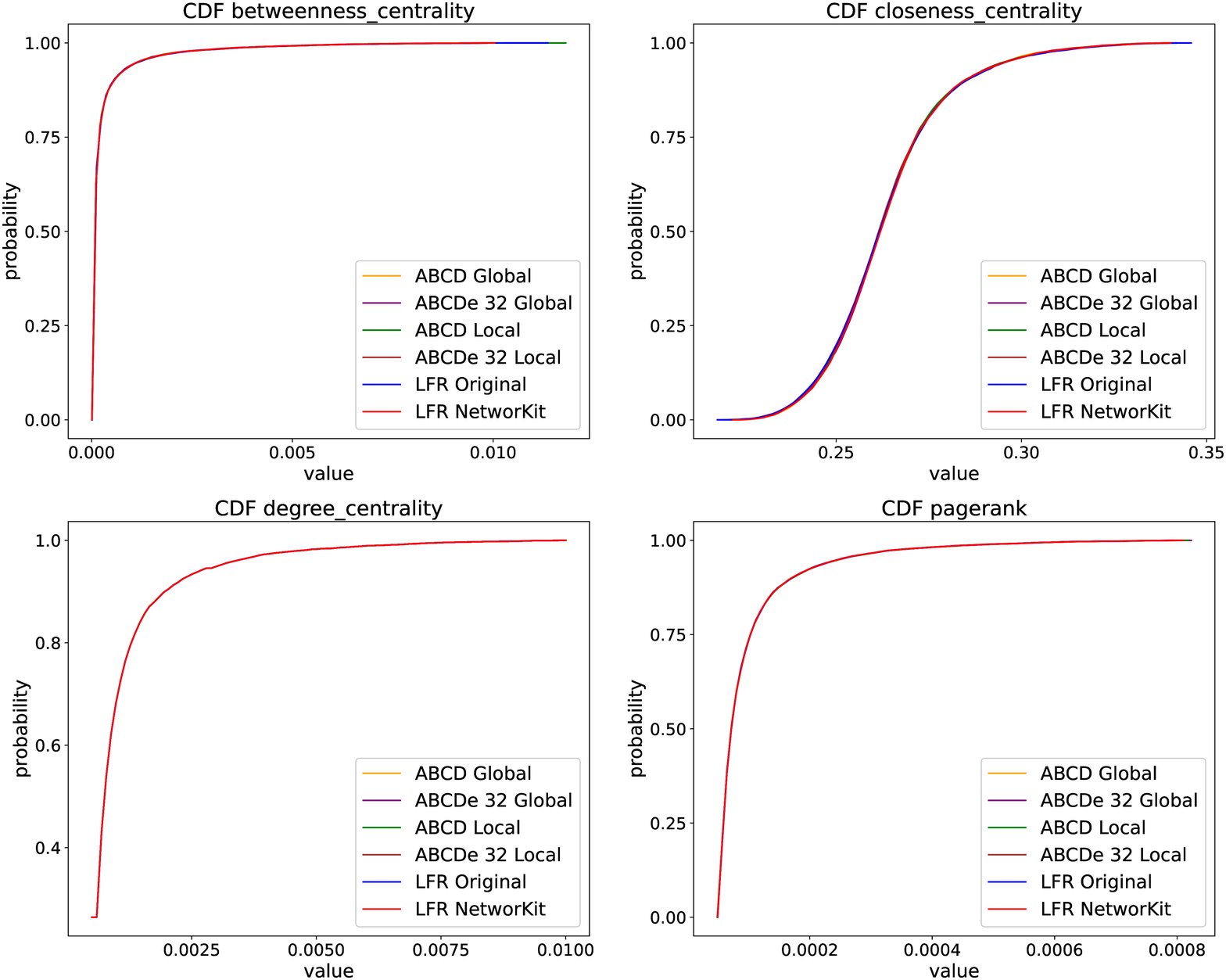}
\caption{Comparison of the distribution of four commonly used centrality coefficients.}\label{fig:cent}
\end{figure}

An important property of a node of a graph is how central it is with respect to the entire graph. Such measures often reflect the relative importance of a node. There are several ways to measure centrality. In Figure~\ref{fig:cent}, we compare the distribution of four commonly used node centrality coefficients: betweenness~\cite{betweenness}, closeness~\cite{closeness}, PageRank~\cite{pagerank}, and degree centrality. As usual, we only provide intuition behind these coefficient and direct the reader to Chapter~3 of~\cite{kaminski2021mining} for more details.
The betweenness centrality for a given node is proportional to the number of shortest paths that pass through that node. For example, in a telecommunications network, a node with higher betweenness would have more control over the network since more information passes through that node.
The closeness centrality is defined as the reciprocal of the sum of the length of the shortest paths between a given node and all other nodes in the graph (assuming the graphs is connected). As a result, nodes that are close to all other nodes are considered more central.
The PageRank centrality, developed by the Google founders, measures the importance of nodes by assuming that important nodes are those that have many important neighbours, something that the degree centrality ignores as it only considers the number of neighbours to score the nodes.
From those plots, we see that all measures are similar for all the benchmark considered.

\subsubsection{Degree Correlation}

In assortative graphs, high degree nodes tend to link to other high degree nodes, while low degree nodes are more often adjacent to low degree nodes. On the other hand, disassortative graph behave differently: high degree nodes tend to be connected to low degree nodes and vice-versa. In order to capture the preference of nodes it is often useful to investigate the degree correlation function which for a given value of $d$ computes the average degree of neighbours of all nodes of degree $d$.
There are two standard ways to measure the overall assortativity of a graph $G$, the degree correlation coefficient and the correlation exponent (see~\cite{corr_fn} and~\cite{corr_coef}, or Chapter~4 in~\cite{kaminski2021mining}).
For both measures, a negative value indicates an assortative network, a value close to zero a neutral one, and a positive value a disassortative one.
Many important properties of graphs, such as speed of spreading information or forming large components, are affected by these measures.

\begin{figure}
\centering
\includegraphics[scale=0.5]{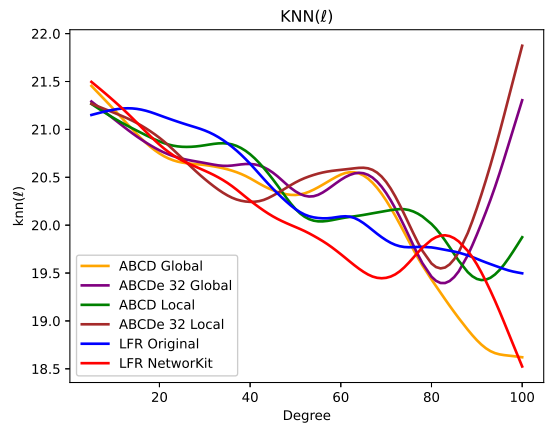}
\caption{Comparison of the degree correlation function.}\label{fig:knn}
\end{figure}

\begin{figure}
\centering
\includegraphics[scale=0.5]{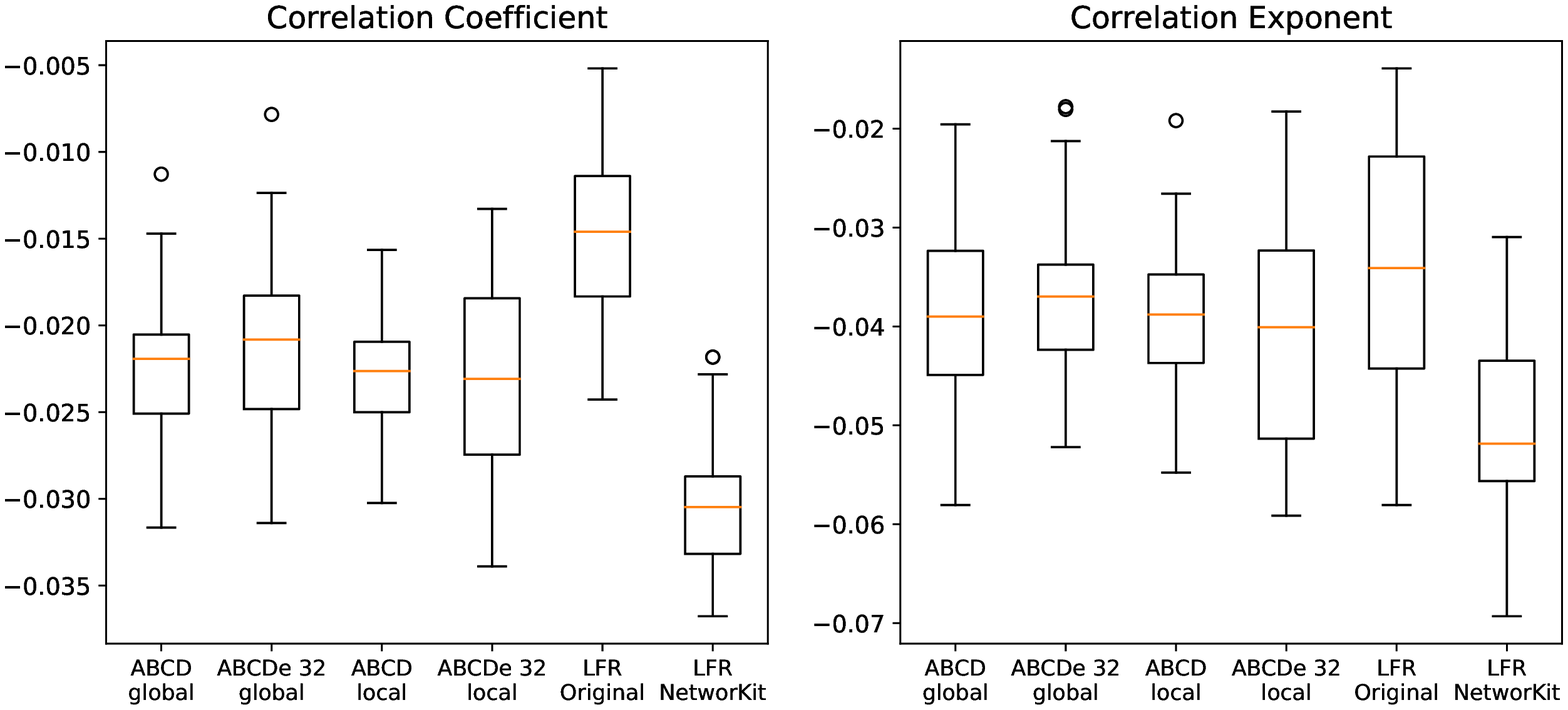}
\caption{Comparison of distribution for the correlation coefficient (left) and the correlation exponent (right).}\label{fig:cor}
\end{figure}

In Figure~\ref{fig:knn}, we see that the shape of the degree correlation function is similar for all random graph models, with a clear negative slope. Hence, all benchmarks produce slightly disassortative graphs.
In Figure~\ref{fig:cor}, we see that the correlation coefficient is the largest for the original \textbf{LFR} and the smallest for \texttt{NetworKit} \textbf{LFR}; both \textbf{ABCD} variants are similar and are roughly in the middle of the range of observed values.
Results for the correlation exponent are similar for both \textbf{ABCD} variants and the original \textbf{LFR}, but it is lower for \texttt{NetworKit} \textbf{LFR}.

\subsubsection{Community Edges and Modularity}

\begin{figure}[ht]
\centering
\includegraphics[scale=0.48]{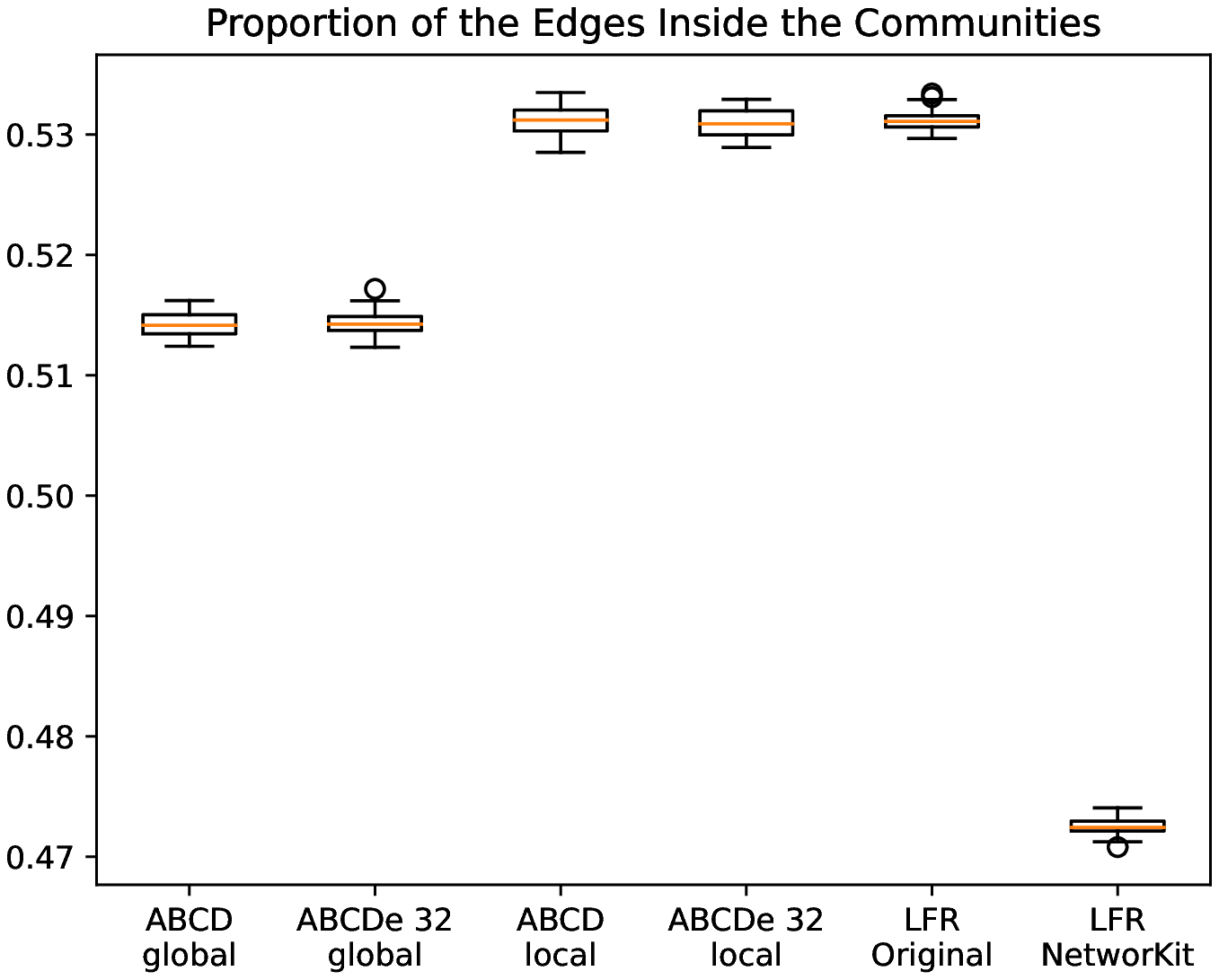}
\includegraphics[scale=0.48]{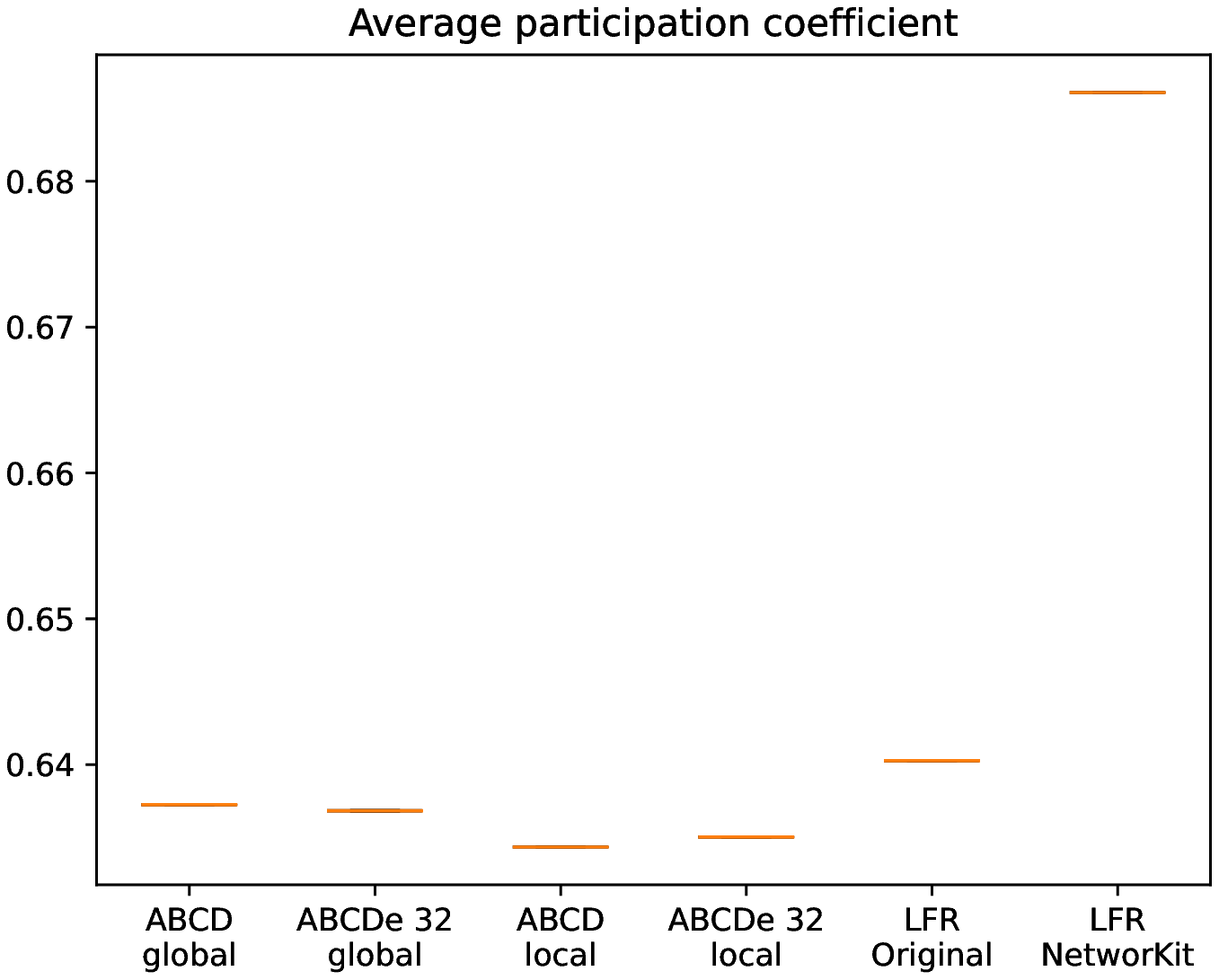}
\caption{Comparison of proportion of edges inside communities (left) and distribution of average participation coefficient (right).}\label{fig:participation}
\end{figure}

We say that an edge is inside a community if both of its nodes are part of the same community; we also refer to those as community edges.
Participation coefficient of a given node, as defined in Chapter~5 of~\cite{kaminski2021mining}, provides more detailed information and measures the distribution of a node's neighbours among the communities of a graph. It is equal to 0 if all of its neighbours are in the same community, and it is close to 1 if its neighbours are equally divided amongst all communities.
In Figure~\ref{fig:participation}, we see that the proportion of edges that are inside communities is very similar for local \textbf{ABCD} and original \textbf{LFR}; it is slightly lower for global \textbf{ABCD} and much lower for \texttt{NetworKit} \textbf{LFR}.
We also plot the average participation coefficient, with the same conclusions (albeit, symmetric).

\begin{figure}
\centering
\includegraphics[scale=0.5]{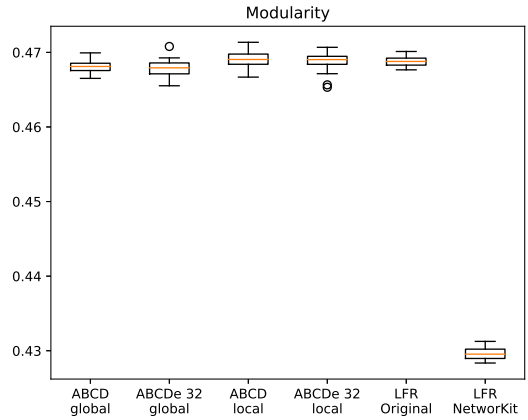}
\caption{Comparison of distribution of modularity.}\label{fig:mod}
\end{figure}

The most important graph property of networks in the context of community detection is the modularity function~\cite{modularity}.
Indeed, the modularity function is often used to measure the presence of community structure in networks. It is also used as a quality function in many community detection algorithms, including the widely used Louvain algorithm.
It is defined as the difference between the proportion of edges inside communities and the expected value of this quantity over some null random graph model. Thus, large (positive) modularity indicates the presence of communities in a graph.
In Figure~\ref{fig:mod}, we see that the modularity function is similar for both \textbf{ABCD} variants and original \textbf{ABCD}, but is again much lower for \texttt{NetworKit} \textbf{LFR}.

\subsubsection{Shortest Paths}

\begin{figure}[ht]
\centering
\includegraphics[scale=0.5]{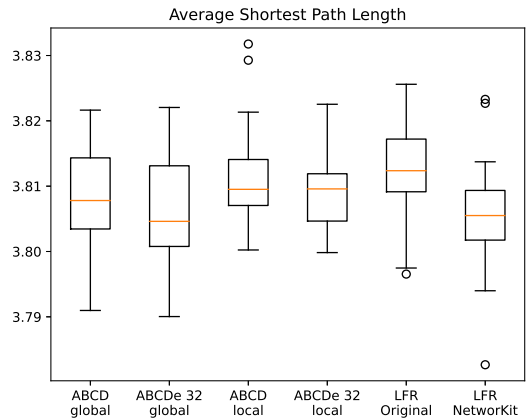}
\caption{Comparison of distribution of average shortest path length.}\label{fig:aspl}
\end{figure}

The shortest path between two nodes in a connected graph is the minimum number of hops to go from one node to the other.
Several studies, inspired by the famous Milgram's small-world experiment, suggest that many real-world networks exhibit surprisingly small average distance between pairs of nodes. As a result, this property (often referred to as ``six degrees of separation'') is often investigated and expected from good random graph models. In Figure~\ref{fig:aspl}, we compare the average shortest path length (obtained via sampling as there are usually too many node pairs to investigate). One can see that this quantity is similar for all benchmark algorithms we tested.

\section{Speed Tests of Graph Generation}
\label{sec:speedtest}

In this section, we present the results of performance comparison of \textbf{ABCD}, \textbf{ABCDe}, and \textbf{LFR}. For \textbf{LFR}, we use the \texttt{NetworKit} implementation as it is approximately 10 times faster than the original \textbf{LFR} implementation.
Moreover, let us mention that both \textbf{ABCDe} and \texttt{NetworKit} \textbf{LFR} implementation support multi-threading. For experiments on these benchmarks, we used 1 to 32 threads.

\subsection{Experiment Setup}

We generated graphs with \texttt{NetworKit} \textbf{LFR} and \textbf{ABCDe} using $k=2^i$ threads for $0 \le i \le 5$. We independently tested single-threaded \textbf{ABCD}.
We compare the time complexity of each method as a function of the number of nodes $n$ in the graph, where $n = 10^i$ for $4 \le i \le 9$. The other parameters were set as for the experiments analyzing properties of the graph presented in the previous section, except for the sizes of the smallest and the largest communities which we vary with $n$, respectively $0.005n$ and $0.2n$.
The range of the values used were chosen based on the objective of the experiment but also the capacity of the machines. For $n < 10^4$ the gain from using multi-threading was negligible whereas experiments with $n > 10^9$ were too computationally expensive for the machines used.

Since the process of generating the background graph is not parallelized, it is expected that the speed of generating \textbf{ABCDe} should depend on the size of the background graph which, in turn, depends on the parameter $\xi$. Results of experiments investigating this relationship (see Figure~\ref{fig:timexiedge}) are presented for a fixed value of $n=10^7$. (As usual, we refer the reader to notebooks for results for other values of $n$.)
In order to couple \textbf{LFR} and global versions of \textbf{ABCD} and \textbf{ABCDe} so that they have a comparable number of edges in the background graph, the mixing parameter $\mu$ was approximated using formula~(\ref{mu_approx}).

The experiments are focused on the graph generation process itself, thus the time spent on generating the degree distribution and community sizes are not included in the comparisons. Similarly to the properties test, pre-generated degree distributions and community sizes were used.

The code for execution and analysis of the experiments was written in Julia 1.6.1 programming language. It is available on GitHub repository\footnote{\url{ https://github.com/bartoszpankratz/ABCDe_Experiments/tree/main/speed\%20test}}, and so are Jupyter notebooks as well as the results for all combinations of parameters\footnote{\url{ https://github.com/bartoszpankratz/ABCDe_Experiments}}.
Julia is a high-level, high-performance, dynamic programming language that recently gains a lot of interest in scientific computing applications ~\cite{juliabezanson}.
\textbf{ABCD} and \textbf{ABCDe} graphs were generated using the following packages written in the Julia language. \texttt{ABCDGraphGenerator v0.1.0}\footnote{\url{ https://github.com/bkamins/ABCDGraphGenerator.jl}} was used for single threaded implementation (\textbf{ABCD}). \texttt{ABCDeGraphGenerator v0.2.4}\footnote{\url{ https://github.com/tolcz/ABCDeGraphGenerator.jl}} was used for a multi threaded implementation (\textbf{ABCDe}).
\textbf{LFR} graphs were generated using Python v3.8.8 with the \texttt{NetworKit v8.1} module\footnote{\url{https://networkit.github.io/}}.
This implementation of the \textbf{LFR} algorithm was chosen because it is considered to be the fastest one available at the time of running of the experiments.

Experiments were performed on the machines with 32 Intel Xeon Processors (Cascadelake) 2.30 GHZ vCPUs with 160GB RAM memory, 120GB disk space and Ubuntu 20.04.1 operating system. They were run simultaneously on four machines for approximately two weeks, totalling in over 2000 vCPU hours.

\subsection{Results}

We first show results of experiments comparing multi-thread \textbf{ABCDe} with single-thread \textbf{LFR}. Then, we compare it with multi-thread \textbf{LFR} for varying $n$ (the number of nodes) and $\xi$ (the mixing parameter), respectively.
As a general conclusion, we see 10--50 fold speed-ups using \textbf{ABCDe} instead of \textbf{LFR}. Another general observation is that using multiple threads does yield speed-ups but the difference could be small beyond 8 threads.

\subsubsection{Speed-up with Multi-thread ABCDe}

In Table~\ref{tab:speed}, we show the speed-up of \textbf{ABCDe} algorithm over \texttt{NetworKit} \textbf{LFR} run on a single thread. In particular, even single-thread \textbf{ABCDe} algorithm is faster than \texttt{NetwrorKit} \textbf{LFR} for varying values of graph size $n$ and $\xi$ parameter; we observe from 13 to 45 times speed-up which is a practically significant improvement regarding that the \textbf{LFR} generation time for graphs with $n=10^9$ nodes was roughly 60 hours in our experiments. Let us also point out that the new \textbf{ABCDe} algorithm (run in a single-thread mode) is faster than the old sequential \textbf{ABCD} implementation by around 30\% (as reported in Figure~\ref{fig:timezsizeedge}) (the changes leading to these speedups are of code optimization nature, like reduction of volume of memory allocations, and were guided by profiling of code runtime; they did not introduce new algorithmic ideas). Additionally, one can observe that increasing the number of threads improves the speed of the \textbf{ABCDe} graph generation, with best results achieved typically when using 8 threads or more (unfortunately the individual timings were not very stable since it was impossible to maintain the same level of system load of the test machine in the cloud infrastructure because the time of running of the whole experiment was long).

\begin{table}
\centering
{\small
\begin{tabular}{r|r|r|r|r|r|r|r}
$\xi$ & $n$ & \textbf{ABCDe1} & \textbf{ABCDe2} & \textbf{ABCDe4} & \textbf{ABCDe8} & \textbf{ABCDe16} & \textbf{ABCDe32} \\
\hline \hline
0.2 & $10^4$ & 32.14 & 45.11 & 57.2 &  {\bf63.73} & 53.98 & 38.58\\
0.2 & $10^5$ & 12.96 & 17.05 & 24.22 & 25.71 &  {\bf28.07} & 26.62\\
0.2 & $10^6$ & 16.96 & 25.28 & 35.82 & 36.01 & 39.66 &  {\bf39.95}\\
0.2 & $10^7$ & 23.52 & 35.47 & 47.21 & 50.73 & 54.19 &  {\bf54.21}\\
0.2 & $10^8$ & 15.31 & 23.79 & 32.57 & 36.5 & {\bf37.24} &  35.77\\
0.2 & $10^9$ & 45.84 & {\bf53.17} & 50.77 & 47.48 & 46.56 &  40.29\\
\hline
0.5 & $10^4$ & 19.95 & 24.92 & 29.21 &  {\bf30.89} & 27.51 & 21.99\\
0.5 & $10^5$ & 20.73 & 29.09 & 34.03 & 33.8 &  {\bf36.17} & 35.55\\
0.5 & $10^6$ & 30.18 & 43.44 & 51.05 & 51.26 & 50.95 &  {\bf56.79}\\
0.5 & $10^7$ & 33.89 & 44.04 & 47.34 & 47.89 & 49.9 &  {\bf50.38}\\
0.5 & $10^8$ & 27.96 & 39.04 & 42.48 & 43.75 &  {\bf44.81} & 44.17\\
0.5 & $10^9$ & 38.19 & 39.96 & {\bf41.52} & 37.91 &  37.31 & 35.06\\
\hline
0.8 & $10^4$ & 29.34 &  33.57 & {\bf35.55} & 35.26 & 34.22 & 29.81\\
0.8 & $10^5$ & 35.76 & 42.93 & 46.36 &  43.89 & 46.71 & {\bf47.51}\\
0.8 & $10^6$ & 43.46 & 57.65 & 56.28 & 57.69 & 61.3 &  {\bf69.14}\\
0.8 & $10^7$ & 39.78 & 43.1 & 46.88 & 47.83 & 46.31 &  {\bf48.35}\\
0.8 & $10^8$ & 38.52 & 44.49 & 46.3 & 48.31 &  {\bf49.1} & 46.57\\
0.8 & $10^9$ & 37.69 & 41.57 & 41.14 & 44.24 &  {\bf46.05} & 34.27\\
\end{tabular}
\vspace{.25cm}
}
\caption{Comparison of graph generation speed in reference to \texttt{NetworKit} \textbf{LFR} (single threaded) as a function of graph size and $\xi$. The values reported are normalized so they reflect how many times faster is the graph generation process in comparison to the reference \texttt{NetworKit} \textbf{LFR} implementation; hence, the larger the values the better. Number \textbf{X} in the name of the model \textbf{ABCDeX} represents the number of threads used to generate the graph. The best results are shown in bold.}
\label{tab:speed}
\end{table}

\subsubsection{Comparing Multi-thread \textbf{ABCDe} and \textbf{LFR}---Varying $n$}

\begin{figure}
\centering
\includegraphics[scale=0.45]{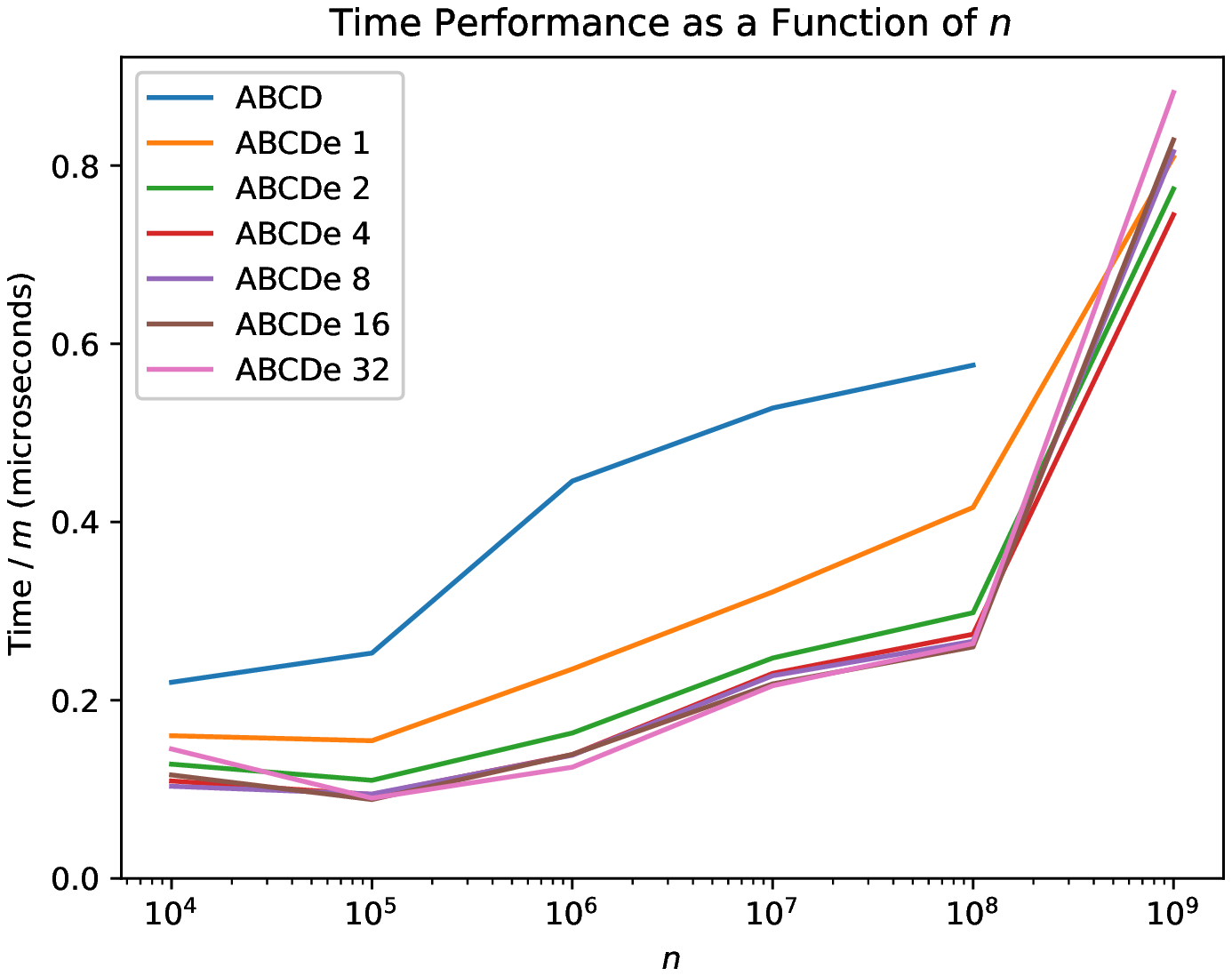}
\includegraphics[scale=0.45]{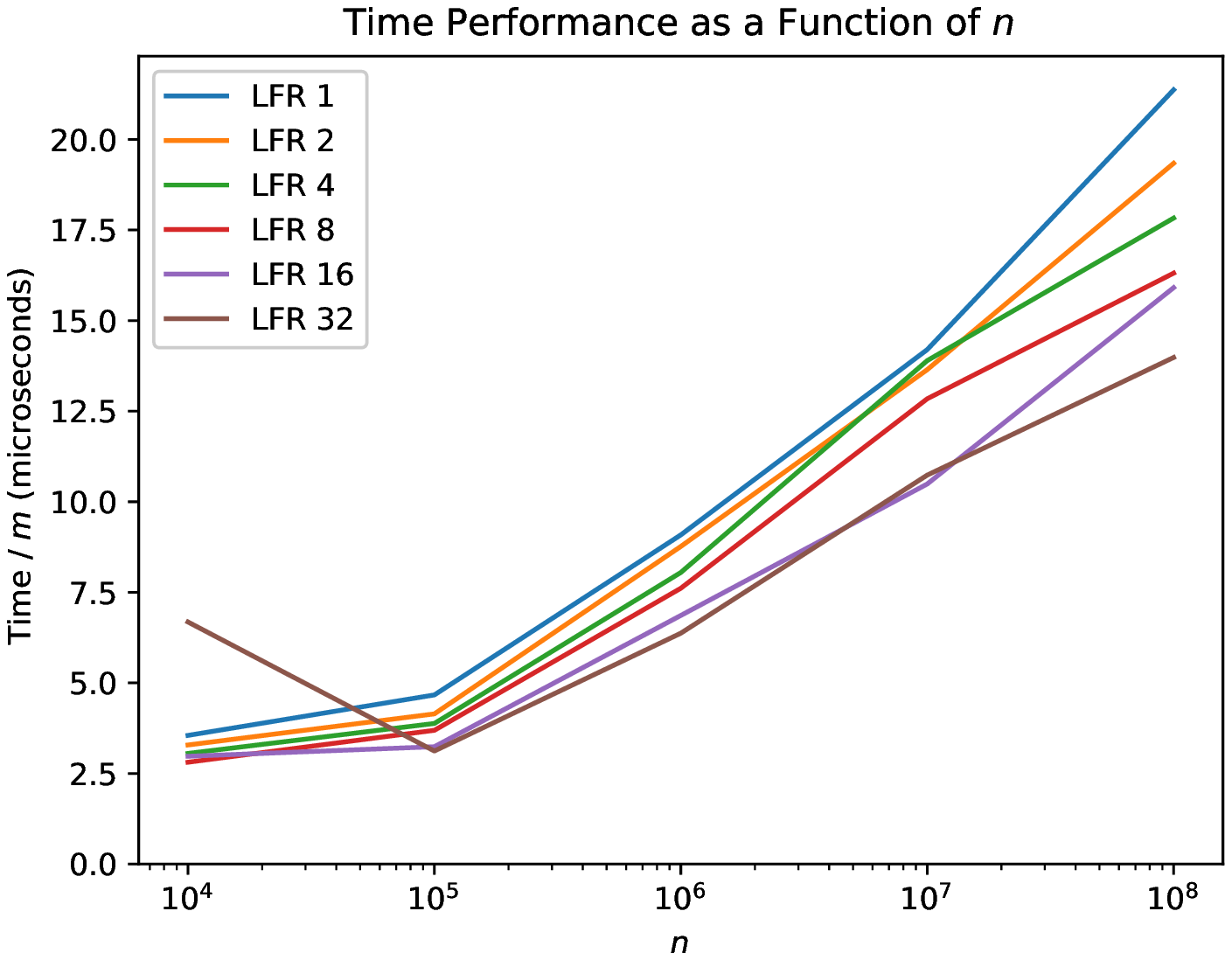}
\caption{Comparison of time (in microseconds) to generate a single edge as a function of the graph order. In the left plot we present statistics for \textbf{ABCD} model and in the right plot for \textbf{LFR} model. Numbers in the legend are the number of threads used to generate the graphs.}
\label{fig:timezsizeedge}
\end{figure}

In Figure~\ref{fig:timezsizeedge}, we present a comparison of average generation time per edge as a function of $n$, the order of a graph, for \textbf{ABCDe} and \texttt{NetworKit} \textbf{LFR}, with varying number of threads (for sequential \textbf{ABCD} and for \textbf{LFR} the experiments were not run for $n=10^9$ as their run-time was very long; we only run the \textbf{LFR} algorithm once to measure its baseline timing for Table~\ref{tab:speed}, which was around 60 hours). The parameter $\xi$ was fixed to $\xi=0.5$.
The results are consistent with the data presented in Table~\ref{tab:speed} with \textbf{ABCDe} over 10\ times faster than \texttt{NetworKit} \textbf{LFR} for all considered scenarios.
However, there are two additional characteristics that should be noted.
First of all, observe that both algorithms decrease speed of edge generation as $n$ increases. However, this speed decrease is much slower for \textbf{ABCDe}. Indeed, in the case of \textbf{ABCDe}, moving from $n=10^4$ to $n=10^8$, the edge generation time increases by no more than a factor of 2.5. On the other hand, for \texttt{NetworKit} \textbf{LFR}, the increase is over 5 fold and so a better scaling is observed for \textbf{ABCDe}. For $n=10^9$ we observe a bump in processing time of \textbf{ABCDe}. This bump is due to the fact that this size of graph was at the limit of processing capability of the available infrastructure. A similar bump for \textbf{LFR} due to the same resource limits reasons is expected, which conformed in Table~\ref{tab:speed}, where the relative speedup of \textbf{ABCDe} over \textbf{LFR} does not drop for $n=10^9$.
The second aspect we would like to highlight are the speed-ups obtained when increasing the number of threads.
Here one can also see that \textbf{ABCDe} scales better; for example, when moving from 1 to 2 threads there is a noticeable speed-up in \textbf{ABCDe} (roughly 20\%), while for \texttt{NetworKit} \textbf{LFR} it is relatively small (less than 5\%).
Moreover, when we compare using 1 vs.\ 32 threads, the speed-up for \textbf{ABCDe} is over 2-fold, while for \texttt{NetworKit} \textbf{LFR} it is less than 1.5-fold.

\subsubsection{Comparing Multi-thread \textbf{ABCDe} and \textbf{LFR}---Varying $\xi$}

\begin{figure}[ht]
\centering
\includegraphics[scale=0.4]{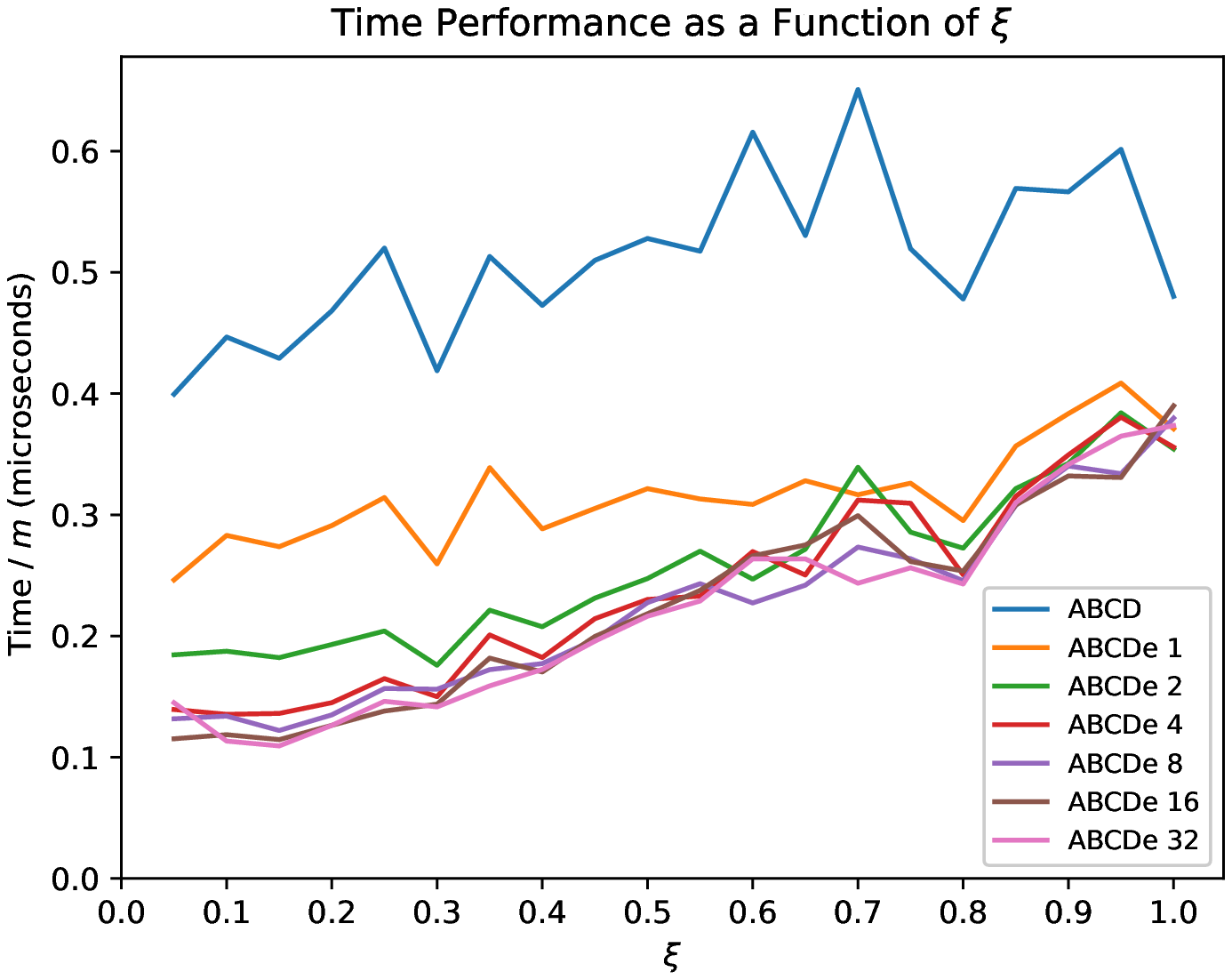}
\includegraphics[scale=0.4]{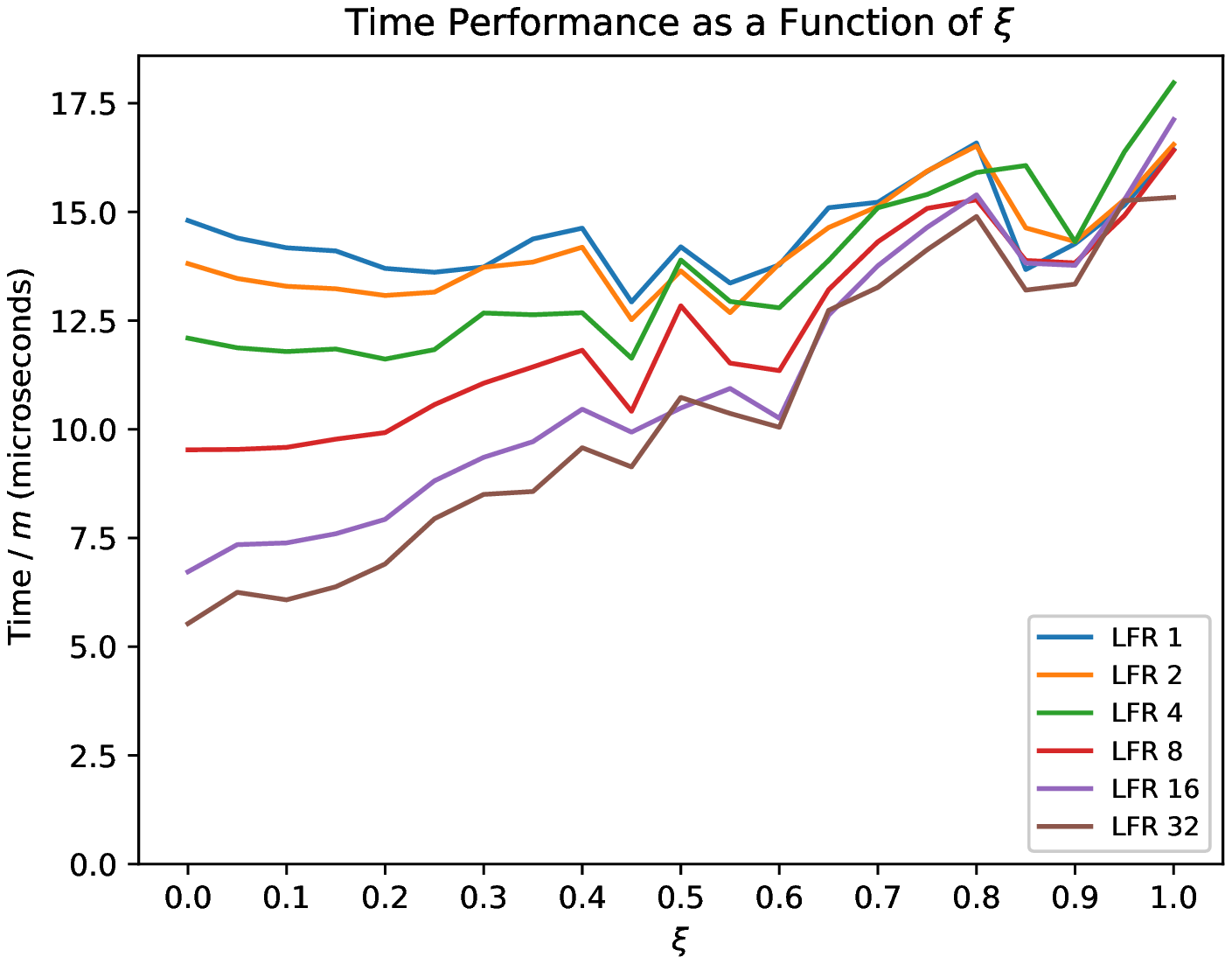}
\caption{Comparison time (in microseconds) to generate a singe edge as a function of the parameter $\xi$. In the left plot we present statistics for \textbf{ABCD} model and in the right plot for \textbf{LFR} model. Numbers in the legend are the number of threads used to generate the graphs.}
\label{fig:timexiedge}
\end{figure}

For the last experiment, we fix the number of nodes to $n=10^7$ and vary the mixing parameters~$\xi$.
The results are shown in Figure~\ref{fig:timexiedge}.
The speed-up obtained with \textbf{ABCDe} is still enormous compared to \textbf{LFR}, but we also see that both algorithms behave similarly with respect to the different number of threads used.
The largest speed-ups with multiple threads are observed for low values of $\xi$; for \textbf{ABCDe} model, this is not too surprising since in that case most of the time is spent for generating community graphs which are completely independent tasks.

\section{Conclusions and Future Directions}

In this paper, we showed that the multi-thread \textbf{ABCDe} is over 10 times faster than \textbf{LFR} and scales better than parallel implementation of the \textbf{LFR} algorithm provided in \texttt{NetworKit}. Moreover, the algorithm is not only faster but graphs generated by \textbf{ABCDe} algorithm have similar properties to graphs generated by the original \textbf{LFR} algorithm, while the parallelized \textbf{NetworKit} implementation of \textbf{LFR} produces graphs that have noticeably different characteristics. In the paper we described the technical approach we have taken to parallelize the \textbf{ABCDe} algorithm. Additionally, the produced algorithm ensures reproducibility of the generated graph independent from task execution ordering or number of threads used.

Despite the fact that \textbf{ABCDe} is already very fast, there are some ways the generation process can be improved. In particular, the current ``bottleneck'' and an area for further research is to design a faster, parallelized implementation of the background graph. This would allow to make a better use of multiple threads in scenarios when the parameter $\xi$ is large and so most of the edges are present in the background graph. Currently, as it is shown in Figure~\ref{fig:timexiedge}, the scalability with number of threads of \textbf{ABCDe} algorithm degrades as $\xi$ increases.

Another further direction worth investigating would be to generalize the \textbf{ABCD}/\textbf{ABCDe} model to include more sophisticated, higher-order structures as well as to capture the dynamics of networks. A good starting point would be to deal with hypergraphs in which edges (called hyperedges) may contain more than two nodes. Indeed, the modularity function for graphs was recently generalized to hypergraphs~\cite{hypergraphs1} and a number of research groups started working on scalable algorithms~\cite{kumar,hypergraphs2} as well as software implementations such as the \texttt{HyperNetX} package\footnote{\url{https://github.com/pnnl/HyperNetX}}
but there is need for synthetic hypergraph benchmarks. One of the very first attempts include the hypergraph stochastic block model~\cite{HSBM} but now it is time for more realistic models producing power-law degree distribution and other desired properties.
Temporal graph generators that control both the evolution of the degree distribution as well as the distribution of community sizes are even more challenging.
One of the very first such models is \textbf{RTGEN}, A \textbf{R}elative \textbf{T}emporal Graph \textbf{GEN}erator~\cite{RTGEN} that was recently introduced.

\section{Acknowledgment}

Hardware used for the computations was provided by the SOSCIP consortium\footnote{\url{https://www.soscip.org/}}. Launched in 2012, the SOSCIP consortium is a collaboration between Ontario’s research-intensive post-secondary institutions and small- and medium-sized enterprises (SMEs) across the province. Working
together with the partners, SOSCIP is driving the uptake of AI and data science solutions and enabling the development of a knowledge-based and innovative economy in Ontario by supporting technical skill development and delivering high-quality outcomes. SOSCIP supports industrial-academic collaborative research projects through partnership-building services and access to leading-edge advanced computing platforms, fuelling innovation across every sector of Ontario’s economy.

\end{document}